\documentclass[aip,a4paper,reprint,groupedaddress]{revtex4-1}

\usepackage{graphicx}
\usepackage{siunitx}
\usepackage[hyperindex,breaklinks]{hyperref}

\hypersetup{
	colorlinks   = true,
	urlcolor     = blue,
	linkcolor    = blue,
	citecolor   = blue
}

\DeclareSIUnit\Molar{\textsc{m}}
\newcommand{\M}[1]{\SI{#1}{\Molar}}
\newcommand{\mM}[1]{\SI{#1}{\milli\Molar}}
\newcommand{\uM}[1]{\SI{#1}{\micro\Molar}}
\newcommand{\nM}[1]{\SI{#1}{\nano\Molar}}
\newcommand{\nm}[1]{\SI{#1}{\nano \meter}}

\newcommand{\mV}[1]{\SI{#1}{\milli \volt}}
\newcommand{\kHz}[1]{\SI{#1}{\kilo \hertz}}

\begin{document}
\title{Promoting single-file DNA translocations through nanopores using electroosmotic flow}

\author{Niklas Ermann}
\author{Nikita Hanikel}
\author{Vivian Wang}
\author{Kaikai Chen}
\author{Ulrich F. Keyser}
\email{ufk20@cam.ac.uk}
\affiliation{Cavendish Laboratory, University of Cambridge, 19 JJ Thomson Avenue, Cambridge CB3 0HE, UK}

\begin{abstract}
	Double-stranded DNA translocates through sufficiently large nanopores either in a linear, single-file fashion or in a folded hairpin conformation when captured somewhere along its length. We show that the folding state of DNA can be controlled by changing the electrolyte concentration, pH and PEG content of the measurement buffer. At \M{1} LiCl or \M{0.35} KCl in neutral pH, single-file translocations make up more than $90\%$ of the total. We attribute the effect to the onset of electroosmotic flow from the pore at low ionic strength. Our hypothesis on the critical role of flows is supported by the preferred orientation of entry of a strand that has been folded into a multi-helix structure at one end. Control over DNA folding is critical for nanopore sensing approaches that use modifications along a DNA strand and the associated secondary current drops to encode information.
\end{abstract}


\maketitle

\section{Introduction}

Solid-state nanopores are finding increasingly widespread use in the study and sensing of single molecules \cite{Dekker2007, Muthukumar2015, Shi2017}. A novel class of sensing approaches uses modifications along the length of double-stranded DNA to encode information. Such modifications cause secondary drops in the already reduced current level during translocation and can be associated with the presence of a target molecule, the outcome of a displacement reaction or a certain base sequence along the DNA strand \cite{Bell2016, Sze2017, Beamish2017, Kong2017, Singer2010}. 

\begin{figure*}
	\includegraphics{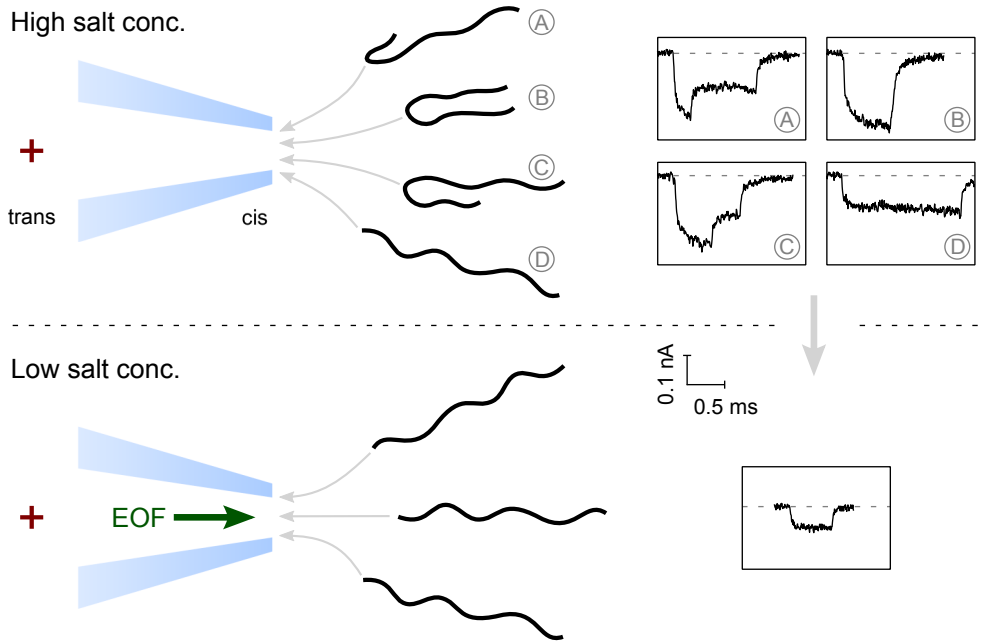}
	\caption{Controlling DNA folding during translocation through a nanopore. At high ionic strength (top), DNA is frequently captured along its contour, causing a part of the strand to fold back onto itself which results in multiple current levels. We expect that electroosmotic flow (EOF) from the pore at low salt concentrations leads to a larger share of single-file translocations (bottom). The right part of the figure shows current traces corresponding to different folding states of 8000 base pair double-stranded DNA at \M{4} (top) and \M{1} LiCl (bottom).
	}
	\label{fig:overview}
\end{figure*}

Common to these approaches is the need for reliable readout of the secondary current drops during DNA translocation. This is facilitated by single-file entry into the pore where DNA is captured at one of its ends. However, hairpin conformations in which a part of the strand folds back onto itself have been reported in a variety of pore types \cite{Li2003, Storm2005, Mihovilovic2013, Steinbock2010}. The top part of figure \ref{fig:overview} shows how DNA capture along its contour causes translocation in a folded state which results in an additional level in the current signature. The presence of two double strands in the pore complicates the identification of secondary current modulations in the folded part of the translocation. While small diameter pores alleviate the problem by forcing single-file entry, they suffer from other issues such as lower capture rates and often non-specific interactions with the pore walls. It is therefore desirable to enhance single-file DNA translocations through other means.

In this work we show that in nanopores fabricated from glass capillaries, the folding state of double-stranded DNA can be controlled through the electrolyte concentration of the measurement solution. We hypothesize that the share of single-file translocations increases in the low salt regime due to the onset of electroosmotic flow (EOF) from the pore.

\section{Methods}
\label{sec:methods}

The nanopores in this work were fabricated by laser-assisted pulling of quartz capillaries as described previously\cite{Bell2015}. This resulted in conical pores with diameters of \nm{12.5} $\pm$ \nm{2.8} as calculated from current-voltage curves and a double-taper resistance model \cite{Bell2015}. Electrolyte solutions were prepared from lithium and potassium chloride (LiCl, KCl) in their powdered form. pH values of \SI{5.5}{}, \SI{7}{} and \SI{8}{} resulted from buffering with \mM{10} MES, \mM{10} HEPES or \mM{10} Tris/\mM{1} EDTA respectively. For a number of measurements we added polyethylene glycol (PEG) with a molecular weight of \SI{8000}{\gram\per\mole} at a concentration of \uM{75}. All salts, buffers and polymers were purchased from Sigma Aldrich.

Experiments investigating DNA folding were carried out with commercially available 8000 base pair double-stranded DNA (NoLimits, Thermo Fisher Scientific, MA, US). Using DNA nanotechnology, we also added end modifications to a previously described 7228 base pair scaffold\cite{Bell2015} to explore the effect of asymmetry on translocating DNA. Staple strands were designed to fold one end of the scaffold back onto itself, creating a 4-helix bundle consisting of 4 double strands. More detail can be found in section 1 of the supplementary information (SI). All DNA samples were added to the \textit{cis} reservoir at concentrations between \nM{0.5} and \nM{5}, see figure \ref{fig:overview} for definitions of the \textit{cis} and \textit{trans} reservoirs. For the study of fluid flows we used \mM{1} of uncharged dextran with a molecular weight of \SI{6000}{\gram\per\mole}.

Current measurements were carried out using an Axopatch 200B patch-clamp amplifier (Molecular Devices, CA, US) with an internal filter setting of \kHz{100} at a bias voltage of \mV{600}. An external 8-pole analog low-pass Bessel filter with a cut-off frequency of \kHz{50} (900CT, Frequency Devices, IL, US or 3382, Krohn-Hite, MA, US) further reduced the noise before recording with a data acquisition card (PCIe-6351, National Instruments, TX, US) and a custom-written LabView program at \kHz{250}. 

Translocation events were identified as deviations from the baseline current greater than a set threshold. For end-modified DNA molecules, a peak finding algorithm adapted from a previous publication\cite{Plesa2015a} then detected the secondary current drops produced by the multi-helix end. We detected DNA folding states with the \textit{MultiNest} Bayesian inference algorithm, which estimates the parameters of two models describing a folded and unfolded translocation respectively \cite{Feroz2009, Buchner2014}. Our approach allows comparison of the two models purely based on their relative Bayesian evidence values without further user-defined thresholds. The position of previously identified secondary current drops relative to the fitted event shape then defined the orientation of entry for asymmetric DNA strands. Section 2 of the SI provides further detail on the data analysis procedure.

\section{Results}

\begin{figure*}
	\includegraphics{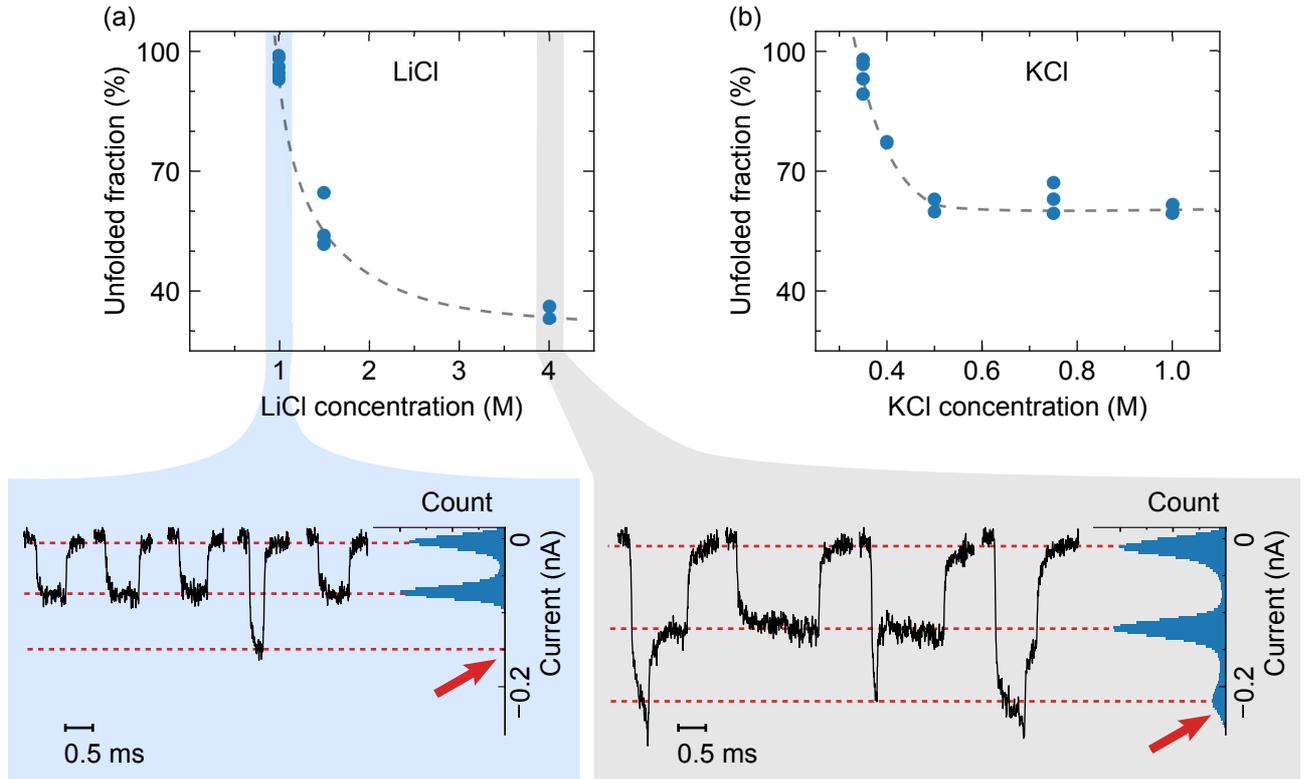}
	\caption{The share of single-file (unfolded) events can be controlled by tuning the electrolyte concentration. \textbf{(a)} Share of unfolded translocations as a function of LiCl concentration at pH 8. Each symbol shows a measurement from a different nanopore with $>$ 500 events. From \M{4} to \M{1}, the proportion of unfolded events increases from below $40\%$ to above $90\%$. The insets in the lower part of the figure show sample traces at \M{1} and \M{4} LiCl respectively, as well as current histograms of the concatenated event data for single nanopore measurements. The third peak in the histograms corresponding to a folded part of the translocation disappears almost completely in \M{1} LiCl (red arrows). \textbf{(b)} Same plot as in (a) for KCl at pH 8. The same effect is observed, with the share of unfolded events starting to increase at a lower electrolyte concentration of \M{0.5}. Dashed lines have been added as guides to the eye in both plots.}
	\label{fig:folding}
\end{figure*}

As outlined in figure \ref{fig:overview}, DNA entering a nanopore can do so linearly or in a folded conformation if captured somewhere along its length. We investigated the relative frequency of both types of translocations in varying electrolyte concentrations and buffer conditions. Figure \ref{fig:folding}a shows the share of single-file, unfolded translocations of 8000 base pair double-stranded DNA as a function of LiCl concentration. Each symbol represents a measurement on a separate nanocapillary with $>$ 500 events. The fraction of single-file translocations increases significantly at low ionic strength from below $40\%$ to above $90\%$. This is reflected in the current histograms in the lower part of figure \ref{fig:folding}, which show a negligible current level corresponding to folded translocations at low salt concentrations as indicated by the red arrows. To rule out a salt-specific effect we carried out the same measurement for KCl. Figure \ref{fig:folding}b again shows the share of unfolded events, this time as a function of KCl concentration. The same increase in single-file translocations can be observed as the ionic strength is lowered. 

Since the persistence length of DNA saturates at the electrolyte concentrations used here, the unfolding effect at low ionic strength cannot be attributed to a change in the energy required to bend the strand \cite{Baumann1997}. An alternative explanation could be increased electrostatic repulsion between the negatively charged DNA and pore walls as the salt concentration is lowered. However, this also seems improbable given a Debye length of less than \nm{0.5} for the electrolyte concentrations shown here and pore diameters around \nm{12.5}. A more likely explanation lies in the emergence of electroosmotic flow (EOF) from the pore as the ionic strength is lowered. Electroosmotic phenomena on the nanoscale have been studied extensively and are well-described by the Debye-H\"{u}ckel theory, which predicts an exponentially decaying surplus of counterions away from the charged capillary surface in electrolyte solution \cite{Schoch2008}. Due to the accumulation of net charge, an electric field applied along the surface exerts a force on the fluid setting it in motion. For an infinite charged cylinder, the longitudinal flow velocity $v$ can be shown to scale as 

\begin{eqnarray}
\label{eq:vel}
v \propto \sigma \lambda \propto \sigma c^{-\frac{1}{2}}
\end{eqnarray}

\noindent
where $\sigma$ is the surface charge, $\lambda$ the Debye length and $c$ the electrolyte concentration \cite{Wong2007}. Equation \ref{eq:vel} shows that the flow velocity scales inversely with the square root of the electrolyte concentration, meaning that the effect becomes more pronounced in aqueous solutions at low salt concentrations. We therefore hypothesized that the increased share in unfolded events could be caused by fluid flow from the pore.

\begin{figure}
	\includegraphics{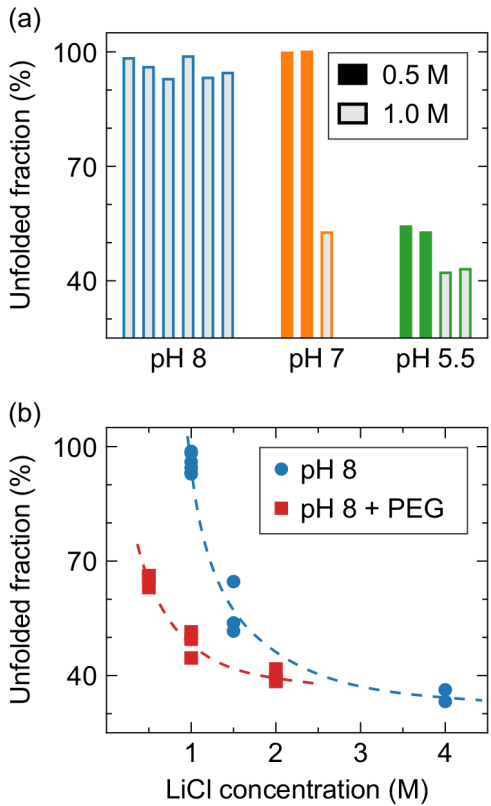}
	\caption{Electroosmotic flow (EOF) controls the fraction of unfolded events. \textbf{(a)} Share of single-file translocations at fixed concentrations of \M{1} and \M{0.5} LiCl for different pH values. Each bar represents a measurement on one nanopore. At \M{1} (empty bars), quenching EOF by lowering pH decreases the fraction of unfolded events. Reducing the electrolyte concentration to \M{0.5} (filled bars) recovers the high share of single-file translocations, albeit at a lower pH of 7, while pH 5.5 still induces folding and no translocations are observed at pH 8. \textbf{(b)} Share of unfolded events as a function of LiCl concentration for buffer containing polyethylene glycol (PEG). Blue circles reproduce the data in figure \ref{fig:folding}a for reference, red squares show measurements with \uM{75} PEG 8000 added to the buffer solution. Quenching EOF through the addition of PEG reintroduces DNA folding at low ionic strength. Dashed lines serve as guides to the eye.}
	\label{fig:eoffolding}
\end{figure}

We tested our explanation by modifying EOF at a given salt concentration. The negative surface charge of glass is strongly pH dependent as silanol groups remain protonated in acidic conditions \cite{Behrens2001}. Lowering the pH thus removes some of the charge, which reduces EOF. Figure \ref{fig:eoffolding}a shows the share of single-file events in varying pH at fixed LiCl concentrations of \M{0.5} and \M{1.0}. At \M{1} LiCl (empty bars), increasing the acidity to pH 7 and eventually 5.5 progressively reduces the fraction of unfolded translocations, as expected from a reduction in EOF. Strikingly, for \M{0.5} LiCl (filled bars) the high levels of unfolded events now appear at pH 7. pH 5.5 again reduces the share of single-file translocations, albeit not to the same extent as in \M{1} LiCl, while an extremely low capture rate at pH 8 precluded the collection of sufficient data. This shows that DNA folding can be carefully tuned by changing the electrolyte concentration and acidity of the measurement solution. The changes in the share of unfolded events correlate with the changes in EOF, pointing to fluid flows as the reason for variations in the DNA folding state.

As mentioned previously, however, an alternative explanation lies in less electrostatic repulsion between pore and DNA due to the reduced surface charge associated with lower pH. We therefore carried out measurements with PEG polymers added to the solution under otherwise identical conditions. PEG adsorbed onto the charged surface has been shown to quench EOF by restricting the mobility of counterions and preventing viscous coupling between the Debye layer and the channel bulk \cite{Doherty2002, Hickey2009}. Importantly, the addition of PEG molecules has the same effect of reintroducing folding at low ionic strength as pH modifications: the red squares in figure \ref{fig:eoffolding}b show that buffer containing \uM{75} PEG 8000 considerably weakens the increase in the fraction of single-file translocations at low salt concentrations. Since PEG is not expected to significantly change the electrostatics of the translocation process, these results confirm EOF as the origin of changes in the share of unfolded events.

\begin{figure}
	\includegraphics{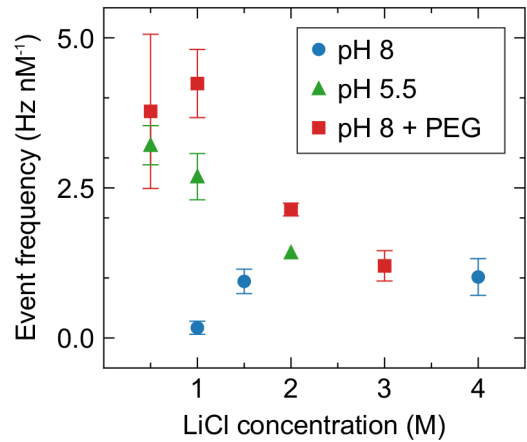}
	\caption{Changes in the event frequency are consistent with the onset of electroosmotic flow (EOF) at low ionic strength and the quenching of EOF through buffer modifications. Shown frequencies are for the same dataset as in figure \ref{fig:eoffolding}. Error bars show the standard deviation of data from multiple nanopores where more than one measurement was carried out. At high ionic strength, the translocation frequency is independent of buffer conditions. Moving towards lower electrolyte concentration, the event frequency drops sharply at pH 8 but increases at pH 5.5 and in solutions containing PEG. This is consistent with the onset of EOF which prevents entry into the pore at pH 8 but is quenched by lower pH and the addition of PEG.}
	\label{fig:caprate}
\end{figure}

Further support for the EOF hypothesis comes from the capture rate of DNA molecules. Lower than expected event frequencies in $Si_{3}N_{4}$ nanopores have been attributed to flows opposing the entry of DNA into the pore \cite{Chen2004a}. This has been confirmed in simulations on nanopores with gate-modulated surface charges \cite{He2011a}. The onset of EOF should therefore correlate with a decrease in the frequency of observed translocations. Figure \ref{fig:caprate} shows the event frequency as a function of LiCl concentration adjusted for the respective DNA concentration. At pH 8 (blue circles), the translocation rate drops considerably at \M{1} LiCl, with almost no detectable translocations below this concentration. At pH 5.5 (green triangles), the event frequency is comparable to that at pH 8 above \M{2} LiCl where EOF is expected to be negligible. Moving to lower electrolyte concentration, however, the frequency in more acidic conditions far outweighs that at neutral pH. A similar enhancement is obtained through the addition of PEG polymers. This further supports our conclusion on the crucial role of EOF, which starts to influence the DNA translocation rate at \M{1} LiCl in neutral pH, but is quenched at low pH and through the addition of PEG polymers.

\begin{figure}
	\includegraphics{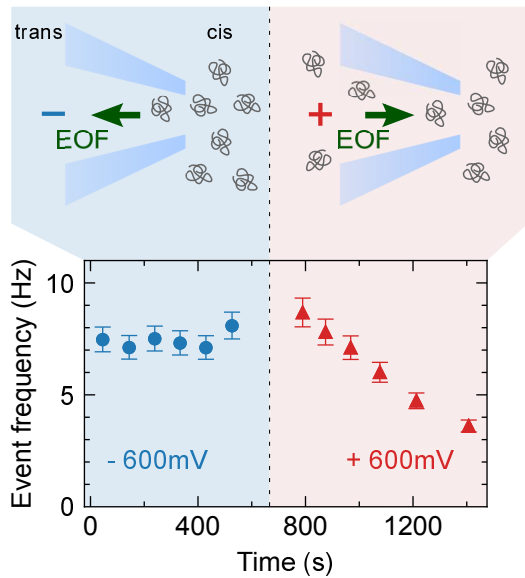}
	\caption{Translocations of uncharged polymers prove the presence electroosmotic flow (EOF) at low electrolyte concentrations. The direction of EOF given the negative surface charge of glass is shown as green arrows in the upper part of the figure. The plot shows the translocation frequency of uncharged dextran 6000 added to the \textit{cis} reservoir at a concentration of \mM{1} in \M{1} LiCl. Each symbol represents the inverse of the mean arrival time of 700 events, while error bars show a $95\%$ confidence interval for the rate parameter of the arrival time exponential distribution. At a voltage bias of -\mV{600} (blue circles), EOF carries a constant number of dextran polymers through the pore to the \textit{trans} side. After inverting EOF with a bias of +\mV{600} (red triangles), previously translocated polymers move back to the \textit{cis} side until none remain in \textit{trans}.}
	\label{fig:dextran}
\end{figure}

While the above observations strongly suggest the presence of EOF at low ionic strength, they only constitute indirect evidence for fluid flows. EOF in nanocapillaries with larger diameters can directly be measured with optical tweezers \cite{Laohakunakorn2013}, but the lower magnitude flows in smaller nanopores at higher electrolyte concentrations require a different approach. In order to demonstrate the presence of EOF, we used uncharged dextran molecules which are driven through the pore by flows instead of the electric field \cite{Wanunu2010}. Figure \ref{fig:dextran} shows the frequency of translocations as a function of time, with \mM{1} dextran 6000 added only to the \textit{cis} reservoir in \M{1} LiCl at pH 8. Initially, \mV{+600} were applied to the \textit{trans} reservoir (not shown in figure \ref{fig:dextran}). No events could be observed, in line with expected EOF from \textit{trans} to \textit{cis} due to the negative glass surface charge. Inverting the bias voltage to \mV{-600} caused a constant rate of translocations over time, as shown by the blue circles in figure \ref{fig:dextran}. In this case, dextran polymers were carried with the flow to the \textit{trans} reservoir, blocking part of the ion current as they passed the pore. Inversion of the voltage back to \mV{+600} led to an initial translocation frequency comparable to that before the switch, followed by a decay over time. The molecules transported to \textit{trans} returned to their original reservoir driven by the inverted EOF, until no molecules remained in \textit{trans}. It should be noted that the detected translocations exhibited a wide distribution of shapes and electronic charge deficits (ECD), indicating varying degrees of dextran aggregation. Nevertheless, the observed events prove the presence of significant EOF in glass nanocapillaries at \M{1} LiCl. 

\begin{figure}
	\includegraphics{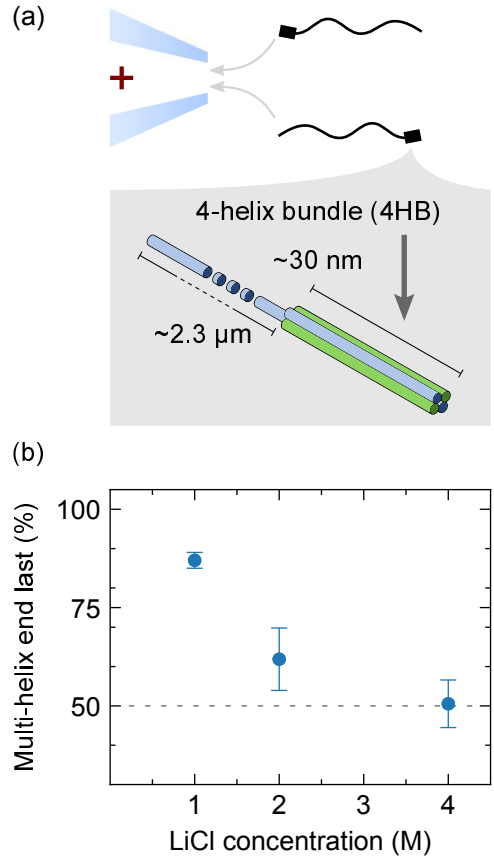}
	\caption{Electroosmotic flow (EOF) influences the orientation of entry for DNA strands with one end shaped into a multi-helix bundle. \textbf{(a)} Design of the modified DNA, each cylinder in the sketch represents one double strand. The end of a single-stranded scaffold is folded into a bundle containing 4 double strands using short staple oligonucleotides, see section 1 of the SI for details. The remainder of the scaffold is hybridized to matching oligonucleotides to create a double strand. \textbf{(b)} Fraction of translocations for which the modified end enters the pore last as a function of LiCl concentration at pH 8 (unfolded events only). Error bars show the standard deviation of measurements on multiple nanopores. At high ionic strength, the DNA is equally likely to get captured at either end. At \M{1} LiCl, a clear preference for entry with the modified end last appears. This is consistent with the onset of EOF from the pore, suggesting the flow biases towards entry of the end with the lower drag force.}
	\label{fig:orient}
\end{figure}

Having demonstrated the presence of considerable fluid flows, we carried out further experiments to investigate how they influence the DNA translocation process. If EOF is responsible for the preferential entry of unfolded DNA, it should also have an effect on the translocation orientation of an asymmetric polymer whose two ends do not experience the same drag force. To test this hypothesis, we designed a modified DNA strand with one of its ends shaped into a bundle of 4 double strands of DNA (4-helix bundle (4HB), see methods section). The other end was left as a blunt-ended double strand. Figure \ref{fig:orient}a shows the design of the asymmetric structure.

As the multi-helix end experiences a higher drag force in fluid flows emanating from the pore, we expect it to enter last when EOF starts to play a role. Indeed, figure \ref{fig:orient}b shows that the share of events where the modified end enters last increases as the ionic strength is lowered. The main rise occurs at \M{1} LiCl, which was identified as the onset of significant EOF in the previous experiments. Fluid flows from the pore therefore create a preference for an orientation of the asymmetric DNA molecule which positions the multi-helix away from the pore entrance. The fraction of unfolded events for the asymmetric DNA structure exhibits the same increase at low ionic strength as the unmodified double strand in figure \ref{fig:folding}, more information can be found in section 3 of the SI.

\section{Discussion \& Conclusion}

We have demonstrated that the folding state as well as orientation of entry of asymmetric DNA strands in glass nanocapillaries depends on the electrolyte concentration of the measurement buffer. Both effects are due to the onset of EOF, which grows in magnitude at low ionic strength. Evidence for significant fluid flows comes from the reappearance of folded events when EOF is quenched by different means, as well as direct observation of flow using uncharged polymers.

The ability to influence the folding state is useful for sensing approaches which rely on single-file translocations to detect secondary current drops produced by modifications or bound targets on the strand. Our results show that by tuning the electrolyte concentration, folded translocations can be almost completely avoided. While this presents a useful alternative to reducing the pore diameter, a number of caveats have to be kept in mind. Firstly, lowering the ionic strength shortens translocation times, which limits the readout resolution for modifications on the strand. The onset of EOF counteracts the speedup, but not sufficiently to prevent an overall increase in DNA velocity. A second caveat concerns the reduced capture rate caused by EOF. If a large enough share of DNA carriers contain the bound target, folded events can simply be discarded while benefiting from the increased translocation frequency at high ionic strength. However, in the case of extremely low analyte concentrations, discarding events may skew the result, particularly if folding is not equally likely for all strand modifications. A large number of unfolded translocations then ensures unbiased results.

To conclude, we have shown that electrolyte concentration-dependent EOF makes it possible to control a DNA strand's conformation and orientation of entry during translocation through a nanopore. Such control is critical for DNA-based sensing techniques that aim to localize modifications along the strand. Our results on glass nanocapillaries are relevant to any system where a significant surface charge induces EOF at low salt concentrations. Promoting single-file entry of DNA may therefore also be achievable in charged 2D membranes \cite{Mao2013}.

\section*{Supplementary material}

See the supplementary material for details on the data analysis procedure and design and folding behavior of the asymmetric DNA structure.

All data accompanying this publication are directly available within the publication.

\begin{acknowledgements}
	N.E. acknowledges funding from the EPSRC, Cambridge Trust and Trinity Hall, Cambridge. N.H. acknowledges the ERASMUS placement organization for ERASMUS+ funding and the German Academic Scholarship Foundation. V.W. is supported by the Winston Churchill Foundation of the United States. K.C. and U.F.K. acknowledge funding from an ERC consolidator grant (Designerpores 647144).
\end{acknowledgements}

\bibliography{lib.bib}

\end{document}